\documentclass[letterpaper,onecolumn,journal,12pt]{IEEEtran}

\usepackage{tabularx,colortbl}
\usepackage{multirow}
\usepackage{dcolumn}
\usepackage{booktabs}
\usepackage{subfig}
\usepackage{cite}
\usepackage[cmex10]{amsmath}
\usepackage{amssymb}
\usepackage{nicefrac}
\usepackage{fixltx2e}
\usepackage{txfonts}
\usepackage{tikz}
\usepackage{pgfplots}
\usetikzlibrary{plotmarks,shapes,positioning,arrows,decorations.markings,fit,calc,patterns}
\usepackage{tikz-timing}
\usetikztiminglibrary[rising arrows]{clockarrows}

\input{register.tikz} % Register shape

\usepackage{ifpdf}
\ifpdf
\usepackage[draft,pdfborder={0 0 0}]{hyperref}
\fi

\begin{document}

\title{A 237 Gbps Unrolled Hardware Polar Decoder}

\author{%
  Pascal Giard,~\IEEEmembership{Student~Member,~IEEE}, %
  Gabi Sarkis, %
  Claude~Thibeault,~\IEEEmembership{Senior~Member,~IEEE}, %
  and Warren~J.~Gross,~\IEEEmembership{Senior~Member,~IEEE}%
  \thanks{G. Sarkis, P. Giard, and W. J. Gross are with the Department of Electrical and Computer Engineering, McGill University, Montr\'eal, Qu\'ebec, Canada (e-mail: \{gabi.sarkis, pascal.giard\}@mail.mcgill.ca, warren.gross@mcgill.ca).}%
  \thanks{C. Thibeault is with the Department of Electrical Engineering, \'Ecole de technologie sup\'erieure, Montr\'eal, Qu\'ebec, Canada (e-mail: claude.thibeault@etsmtl.ca).}}

\maketitle

\begin{abstract}
In this letter we present a new architecture for a polar decoder using a reduced complexity successive cancellation decoding algorithm. This novel fully-unrolled, deeply-pipelined architecture is capable of achieving a coded throughput of over 237 Gbps for a (1024,512) polar code implemented using an FPGA. This decoder is two orders of magnitude faster than state-of-the-art polar decoders.
\end{abstract}

\section{Introduction}
Polar codes provably achieve the symmetric capacity of memoryless channels using the low-complexity successive-cancellation (SC) decoding algorithm \cite{Arikan2009}. However, the SC algorithm is sequential in nature, leading to low-throughput decoders. In \cite{Alamdar-Yazdi2011,Sarkis2014}, new decoding algorithms with the specific aim of reducing the decoding latency and increasing the throughput were proposed.
These algorithms work by decomposing a polar code into its constituent codes and using fast, specialized decoding algorithms on them. They represent polar codes as decoder trees that can be pruned by creating a new node type for each of the recognized constituent code types.

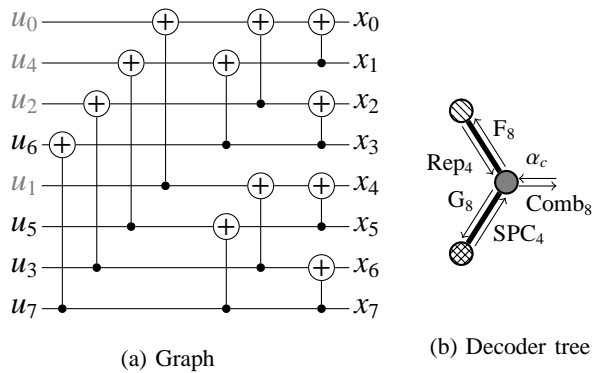
\begin{figure}[ht]
  \centering
  \subfloat[Graph]{\label{fig:sc-graph}\newcommand{\ubit}[1]{$u_{#1}$}
\newcommand{\fbit}[1]{\color{gray}$u_{#1}$}
\newcommand{\ucw}[1]{$x_{#1}$}
\newcommand{\fcw}[1]{\color{gray}$x_{#1}$}
\newcommand{\ub}[1]{$#1$}
\newcommand{\fb}[1]{\color{gray}$#1$}

\begin{tikzpicture}[baseline=(n7s0.center)]

\usetikzlibrary{shapes,positioning,arrows,decorations.markings,fit}

\definecolor{varnode_fill}{RGB}{0,0,0}
\definecolor{chknode_fill}{RGB}{255,255,255}

\tikzset{
  chknode/.style={draw,fill=chknode_fill,circle,minimum size=0.3cm, inner sep=0},
  varnode/.style={draw,fill=varnode_fill,circle,minimum size=0.1cm, inner sep=0},
  sep/.style={rectangle,minimum width=0.25cm, inner sep=0},
  empty/.style={rectangle, inner sep=0},
  bit/.style={circle, inner sep = 0}
}

\matrix[row sep=1mm, column sep=1mm] {
  \node[bit] (n0s0) {\fb{u_0}}; & \node[empty] {}; & \node[empty] {}; &  \node[empty] {}; & \node[chknode] (n0s1) {$+$}; & \node[sep] (s10) {}; & \node[empty] {}; &	\node[chknode] (n0s2) {$+$}; &	\node[sep] (s20) {}; & \node[chknode] (n0s3) {$+$}; && \node[bit] (xn0s4) {\ub{x_0}};\\
  \node[bit] (n4s0) {\fb{u_4}}; & \node[empty] {}; & \node[empty] {}; & \node[chknode] (n4s1) {$+$}; & \node[sep] (s14) {}; & \node[empty] {}; &   \node[chknode] (n4s2) {$+$}; &  \node[sep] (s24) {}; &  \node[empty] {}; & \node[varnode] (n4s3) {};	&& \node[bit] (xn4s4) {\ub{x_1}};\\
  \node[bit] (n2s0) {\fb{u_2}}; & & \node[chknode] (n2s1) {$+$}; & \node[sep] (s12) {}; && &&  \node[varnode] (n2s2) {};    &  \node[sep] (s22) {}; &   \node[chknode] (n2s3) {$+$}; && \node[bit] (xn2s4) {\ub{x_2}};\\
  \node[bit] (n6s0) {\ub{u_6}}; & \node[chknode] (n6s1) {$+$}; & \node[sep] (s16) {}; & \node[empty] {};& \node[empty] {}; & \node[empty] {};&   \node[varnode] (n6s2) {};    &  \node[sep] (s26) {}; & \node[empty] {};&   \node[varnode] (n6s3) {};&& \node[bit] (xn6s4) {\ub{x_3}};\\
  \node[bit] (n1s0) {\fb{u_1}}; &&& & \node[varnode] (n1s1) {};    & \node[sep] (s11) {}; & &  \node[chknode] (n1s2) {$+$}; & \node[sep] (s21) {}; &  \node[chknode] (n1s3) {$+$};&& \node[bit] (xn1s4) {\ub{x_4}};\\
  \node[bit] (n5s0) {\ub{u_5}}; && & \node[varnode] (n5s1) {};     & \node[sep] (s15) {}; & &  \node[chknode] (n5s2) {$+$}; & \node[sep] (s25) {}; &&  \node[varnode] (n5s3) {};&& \node[bit] (xn5s4) {\ub{x_5}};\\
  \node[bit] (n3s0) {\ub{u_3}}; & & \node[varnode] (n3s1) {};    & \node[sep] (s13) {}; &&& &  \node[varnode] (n3s2) {};    & \node[sep] (s23) {}; &     \node[chknode] (n3s3) {$+$};&& \node[bit] (xn3s4) {\ub{x_6}};\\
  \node[bit] (n7s0) {\ub{u_7}}; & \node[varnode] (n7s1) {};    & \node[sep] (s17) {}; &&& &  \node[varnode] (n7s2) {};    && \node[sep] (s27) {}; &    \node[varnode] (n7s3) {};&& \node[bit] (xn7s4) {\ub{x_7}};\\
};
\path[-] (n0s0) edge (n0s1) (n0s1) edge (n0s2) (n0s2) edge (n0s3) (n0s3) edge (xn0s4);
\path[-] (n1s0) edge (n1s1) (n1s1) edge (n1s2) (n1s2) edge (n1s3) (n1s3) edge (xn1s4);
\path[-] (n2s0) edge (n2s1) (n2s1) edge (n2s2) (n2s2) edge (n2s3) (n2s3) edge (xn2s4);
\path[-] (n3s0) edge (n3s1) (n3s1) edge (n3s2) (n3s2) edge (n3s3) (n3s3) edge (xn3s4);
\path[-] (n4s0) edge (n4s1) (n4s1) edge (n4s2) (n4s2) edge (n4s3) (n4s3) edge (xn4s4);
\path[-] (n5s0) edge (n5s1) (n5s1) edge (n5s2) (n5s2) edge (n5s3) (n5s3) edge (xn5s4);
\path[-] (n6s0) edge (n6s1) (n6s1) edge (n6s2) (n6s2) edge (n6s3) (n6s3) edge (xn6s4);
\path[-] (n7s0) edge (n7s1) (n7s1) edge (n7s2) (n7s2) edge (n7s3) (n7s3) edge (xn7s4);

\path[-] (n0s1) edge (n1s1);
\path[-] (n2s1) edge (n3s1);
\path[-] (n4s1) edge (n5s1);
\path[-] (n6s1) edge (n7s1);

\path[-] (n0s2) edge (n2s2);
\path[-] (n1s2) edge (n3s2);
\path[-] (n4s2) edge (n6s2);
\path[-] (n5s2) edge (n7s2);

\path[-] (n0s3) edge (n4s3);
\path[-] (n1s3) edge (n5s3);
\path[-] (n2s3) edge (n6s3);
\path[-] (n3s3) edge (n7s3);

\end{tikzpicture}}
  \subfloat[Decoder tree]{\label{fig:fast-ssc-tree}\definecolor{deepgreen}{RGB}{8, 130, 25}

\begin{tikzpicture}[baseline=(base),
        level/.style={level distance = 6mm},
        level 1/.style={sibling distance=19mm, edge from parent/.style={draw,black,line width=2pt}},
        level 2/.style={sibling distance=9mm, edge from parent/.style={draw,black,line width=1pt}},
        level 3/.style={sibling distance=4mm, edge from parent/.style={draw,black,line width=0.5pt}},
        ]

\tikzset{
frozen/.style={thick,draw=black,fill=white,minimum size=3mm,circle, inner sep=0},
fullspace/.style={thick,draw=black,fill=black,minimum size=3mm,circle, inner sep = 0},
mixed/.style={thick,draw=black,fill=gray,minimum size=3mm,circle, inner sep = 0},
rep_mixed/.style={thick,draw=black,pattern=north west lines,minimum size=3mm,circle, inner sep = 0},
spc_mixed/.style={thick,draw=black,pattern=crosshatch,minimum size=3mm,circle, inner sep = 0},
repspc/.style={thick,draw=black,pattern=vertical lines,pattern color=blue,minimum size=3mm,circle, inner sep = 0}
}

\tikzset{
parallel segment/.style={
   segment distance/.store in=\segDistance,
   segment pos/.store in=\segPos,
   segment length/.store in=\segLength,
   to path={
   ($(\tikztostart)!\segPos!(\tikztotarget)!\segLength/2!(\tikztostart)!\segDistance!90:(\tikztotarget)$) -- 
   ($(\tikztostart)!\segPos!(\tikztotarget)!\segLength/2!(\tikztotarget)!\segDistance!-90:(\tikztostart)$)  \tikztonodes
   }, 
   % Default values
   segment pos=.5,
   segment length=4ex,
   segment distance=-1mm,
},
}

\node[mixed] (3_0){} [grow=left]
	child {node[rep_mixed] (2_0){}
	}
	child {node[spc_mixed] (2_1){}
	}
;
	
\draw[->] (3_0) to[parallel segment] node[above right=-1.5mm] {\footnotesize F$_8$} (2_0);
\draw[->] (2_0) to[parallel segment] node[below left=-1.5mm] {\footnotesize Rep$_4$} (3_0);

\draw[->] (3_0) to[parallel segment] node[above left=-1.5mm] {\footnotesize G$_8$} (2_1);
\draw[->] (2_1) to[parallel segment] node[below right=-1.5mm] {\footnotesize SPC$_4$} (3_0);

\draw[<-] ($(3_0.east) + (0mm, 0.5mm)$) -- node[above=0mm] {\footnotesize $\alpha_c$} ($(3_0.east) + (5mm, 0.5mm)$);
\draw[->] ($(3_0.east) + (0mm, -0.5mm)$) -- node[label={[label distance=-3mm]-2mm:{\footnotesize Comb$_8$}}] {} ($(3_0.east) + (5mm, -0.5mm)$);

\node [circle, below= 5.27mm of 2_1.base] (base) {};

\end{tikzpicture}}
  \caption{From a graph to a Fast-SSC decoder tree.}
  \label{fig:graph-to-tree}
\end{figure}

The field-programmable gate-array (FPGA) implementation of the Fast Simplified Successive Cancellation (Fast-SSC) algorithm presented in \cite{Sarkis2014} can achieve an information throughput of 1 Gbps. Fig.~\ref{fig:sc-graph} is the graph representation for an $(8,4)$ polar code where $u_0$, $u_1$, $u_2$ and $u_4$ are frozen bits. Fig.~\ref{fig:fast-ssc-tree} shows the decoder tree corresponding to Fast-SSC decoding of that $(8,4)$ polar code after tree pruning is applied. The arrows indicate the data flow whereas the annotations correspond to the channel values ($\alpha_c$) or functions as defined in the Fast-SSC algorithm \cite{Sarkis2014}. Notably, the striped node corresponds to a Repetition code of length 4 and the cross-hatched one to a single parity check (SPC) code, also of length 4.

Currently, the fastest realization of a decoder for polar codes is the belief-propagation (BP) decoder of \cite{Park2014}, which achieves a coded throughput of 4.68 Gbps (information throughput of 2.34 Gbps) for a (1024, 512) code on a 65 nm CMOS  application-specific integrated-circuit (ASIC) running at 300 MHz.

In spite of these advances, polar decoders remain slow compared to capacity-approaching codes such as low-density parity-check (LDPC) codes, hampering their adoption for high-speed applications. This work addresses this issue by presenting a new decoder architecture that achieves a coded throughput of 237 Gbps (information throughput of 118.5 Gbps) on an FPGA running at 231 MHz for a (1024, 512) polar code.

\section{Architecture}
Most existing polar decoders (e.g. \cite{Raymond2014,Sarkis2014,Park2014}) minimize area and maximize logic utilization by restricting the decoder to decode a single frame. While this approach lowers implementation complexity, it limits decoding throughput. Instead, we propose generating a code-specific unrolled decoder, fully pipelining its execution so that it processes portions of several frames at once, and adding memory registers for the required data persistence.

\begin{figure}[t]
  \centering
  \resizebox{0.8\columnwidth}{!}{
    \begin{tikzpicture}

\tikzset{
branch/.style={fill,shape=circle,minimum size=3pt,inner sep=0pt},
inv/.style={draw,circle,xshift=-0.62mm,fill=white,minimum size=3.5pt,inner sep=0pt},
block/.style={draw, rectangle, minimum height=3em, minimum width=2em},
spc/.style={draw, rectangle, minimum height=3em, minimum width=2em},
rep/.style={draw, rectangle, minimum height=3em, minimum width=2em},
every picture/.style={font issue=\footnotesize},
font issue/.style={execute at begin picture={#1\selectfont}}
}

%\tiny
\fontsize{8}{10}

\node (ac) at (0.7,0) {$\alpha_c$};

\node[shape=reg] at ($(ac)+(0.775,-0.36)$) (REG1) {$\alpha_c$};

\node[shape=reg] at ($(REG1)+(1.5,0)$) (REG2) {$\alpha_c$};
\node[block] at ($(REG2)+(-0.65,-1.4)$) (F1) {F$_8$};
\node[shape=reg] at ($(F1)+(0.65,-0.35)$) (REG4) {$\alpha_1$};

\node[shape=reg] at ($(REG2)+(1.6,0)$) (REG3) {$\alpha_c$};
\node[rep] at ($(REG4)+(0.85,0.368)$) (Rep1) {Rep$_4$};
\node[shape=reg] at ($(Rep1)+(0.75,-0.375)$) (REG5) {$\beta_1$};

\node[block] at ($(REG3)+(0.8,0.11)$) (G1) {G$_8$};
\node[shape=reg] at ($(REG3)+(1.55,0)$) (REG6) {$\alpha_2$};
\node[shape=reg] at ($(REG5)+(1.55,0)$) (REG7) {$\beta_1$};

\node[spc] at ($(REG6)+(0.9,0.35)$) (SPC1) {SPC$_4$};
\node[shape=reg] at ($(SPC1)+(0.8,-0.35)$) (REG8) {$\beta_2$};
\node[shape=reg] at ($(REG7)+(1.7,0)$) (REG9) {$\beta_1$};

\node[block] at ($(REG8)+(1.05,0.11)$) (Comb1) {Comb$_8$};
\node[shape=reg] at ($(REG8)+(1.975,0)$) (REG10) {$\beta_c$};

\node (bc) at ($(REG10)+(0.7,0.36)$) {$\beta_c$};

\draw[-] (ac.east) -- (REG1.D);

\draw[dotted] ($(REG1.Q)+(0.25,0.75)$) -- ($(REG1.Q)+(0.25,-3.25)$);

\draw[-] (REG1.Q) -- (REG2.D);
\draw[-] (REG1.Q) -- ++(0.15,0) node[branch] {} -- ++(0,-0.45) |- (F1);
\draw[-] (F1) -- (REG4.D);

\draw[dotted] (3.35,0.75) -- (3.35,-3.25);

\draw[-] (REG2.Q) -- (REG3.D);
\draw[-] (REG4.Q) -- (Rep1);
\draw[-] (Rep1) -- (REG5.D);

\draw[dotted] (4.9,0.75) -- (4.9,-3.25);

\draw[-] (REG3.Q) |- ($(G1)+(-0.285,0.25)$);
\draw[-] (REG5.Q) -- ++(0.2,0) node[branch] {} |- ($(G1)+(-0.285,-0.25)$);
\draw[-] (REG6.D) -- ++(-0.125,0) |- (G1);
\draw[-] (REG5.Q) -- (REG7.D);

\draw[dotted] ($(REG6.Q)+(0.125,0.75)$) -- ($(REG6.Q)+(0.125,-3.25)$);

\draw[-] (REG6.Q) -- (SPC1);
\draw[-] (SPC1) -- (REG8.D);
\draw[-] (REG7.Q) -- (REG9.D);

\draw[dotted] ($(REG8.Q)+(0.125,0.75)$) -- ($(REG8.Q)+(0.125,-3.25)$);

\draw[-] (REG8.Q) -- ($(Comb1)+(-0.5,0.25)$);
\draw[-] (REG9.Q) -- ++(0.2,0) |- (Comb1.west);
\draw[-] (REG10.D) -- ++(-0.1,0) |- (Comb1.east);

\draw[dotted] ($(REG10.Q)+(0.125,0.75)$) -- ($(REG10.Q)+(0.125,-3.25)$);

\draw[-] (bc.west) -- (REG10.Q);

\end{tikzpicture}
  }
  \caption{Implementation for $(8,4)$ polar code. Clock signal not routed for clarity.}
  \label{fig:impl}
\end{figure}
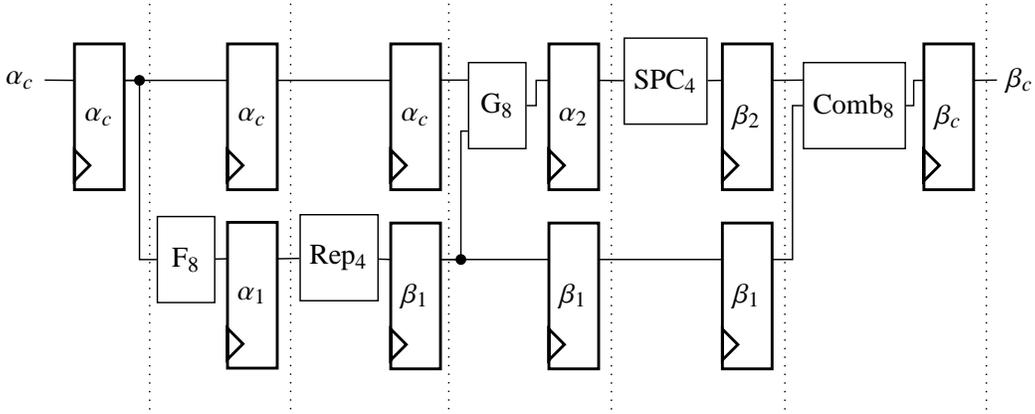

\begin{figure}[t]
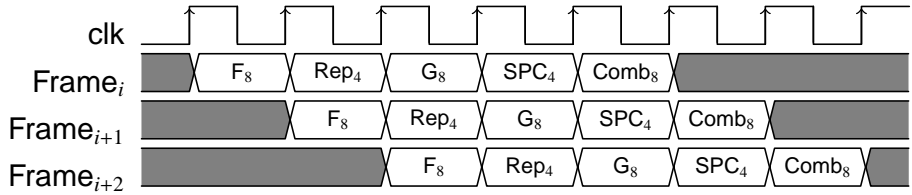

  \centering
  \begin{tikztimingtable}[
    xscale=1.9,yscale=1.5,%
    timing/coldist=0.35,%
    timing/rowdist=1.25,%
    timing/d/text/.append style={scale=1.2},
    semithick
    ]
    clk    & 16{C};\\
    Frame$_i$ & U{}DD{F$_8$}DD{Rep$_4$}DD{G$_8$}DD{SPC$_4$}{DD{Comb$_8$}}UUUUU{}\\
    Frame$_{i+1}$ & UUU{}DD{F$_8$}DD{Rep$_4$}DD{G$_8$}DD{SPC$_4$}DD{Comb$_8$}UUU{}\\
    Frame$_{i+2}$ & UUUUU{}DD{F$_8$}DD{Rep$_4$}DD{G$_8$}DD{SPC$_4$}DD{Comb$_8$}U{}\\
  \end{tikztimingtable}
  \caption{Timing example to decode 3 frames of a $(8,4)$ polar code.}
  \label{fig:unrolled_timing}
\end{figure}

Fig.~\ref{fig:impl} shows the decoder architecture for an $(8, 4)$ polar code. The functional units correspond to the operations shown in Fig.~\ref{fig:fast-ssc-tree}, each of which is followed by a pipeline register to store the operation's output. In addition some pipeline stages do not have any processing logic; they are added to ensure that different messages remain synchronized.
As a result of the pipelined design, at every clock cycle, a frame is output and a new received frame can be loaded as shown in the timing diagram in Fig.~\ref{fig:unrolled_timing}. This deeply-pipelined architecture leads to very high-throughput decoders.

Due to the unrolled nature of the architecture, the growth in resources used is quadratic in code length. It is also affected by the code rate and frozen bit locations as both affect the structure of the decoder tree and, in turn, the number of operations performed in a Fast-SSC decoder.
The amount of memory used is also quadratic in code length and affected by rate and frozen bit locations. In comparison, the Fast-SSC decoder in \cite{Sarkis2014} requires memory that grows linearly in code length.
This growth in resources and memory limits the proposed decoder to codes of moderate lengths when implemented on an FPGA.

\pagebreak

\section{Implementation Results}

The resulting information throughput is $PfR$ bps where $P$ is the width of output bus in bits, $f$ is the execution frequency in Hz and $R$ is the code rate. Latency depends on the frozen bit locations and the constrained maximum width for all modules. In this work, the buses are sized so that all data is transferred simultaneously, i.e. they can carry $N$ log-likelihood ratios (LLRs) and $N$ bit estimates as in \cite{Park2014,Dizdar2014}.

A decoder utilizing the proposed architecture was implemented for a $(1024, 512)$ polar code on an Altera Stratix IV EP4SGX530KH40C2 FPGA. The specialized decoders for repetition and SPC codes were limited to constituent codes of length $\leq 4$, all others were limited a maximum of 1024. Table~\ref{tab:unrolled_results} presents results for two different execution frequencies. It can be observed that, at the cost of some register duplication, the coded (information) throughput can be increased from 210~Gbps (105~Gbps) to 237~Gbps (118.5~Gbps). The latency also decreases from 2.7~$\mu$s to 2.4~$\mu$s at 231 MHz. It can also be noted that, in both cases, register chains are implemented using SRAM blocks.

\begin{table}[h]
  \centering
  \caption{Post-fitting results for a $(1024,512)$ polar code on the Altera\\ Stratix IV EP4SGX530KH40C2 FPGA.}
  \begin{tabular}{c c c c c c}
    \toprule
    \multicolumn{1}{c}{\multirow{2}{*}{LUTs}} & \multicolumn{1}{c}{\multirow{2}{*}{Registers}} & \multicolumn{1}{c}{RAM} & $f$ & Info. T/P & Latency\\
    & & \multicolumn{1}{c}{(bits)} & (MHz) & (Gbps) & (CC)\\
    \midrule
    156,450 & 152,124 & 285,120 & 206 & 105.3 & 559\\
    155,858 & 158,185 & 285,120 & 231 & 118.5 & 559\\
    \bottomrule
  \end{tabular}
  \label{tab:unrolled_results}
\end{table}

Table~\ref{tab:cmp_results} compares the proposed decoder with others from the literature. Notably, the unrolled decoder has 50.7 times the throughput of the BP decoder of \cite{Park2014}, with the latter implemented as a 65 nm CMOS ASIC clocked at 300 MHz. With its maximum of 15 iterations, the BP decoder has a latency that is 21 times higher than the proposed decoder. The Altera Stratix IV FPGA is built using the more recent 40 nm technology. The delay gain between 65 nm and 40 nm CMOS technology is little over 1.23 as this corresponds to the gain between 65 nm and 45 nm \cite{Wong2011}. However, the speed gain of building an ASIC instead of using an FPGA was shown to be from 3.4 to 4.6 \cite{Kuon2007}.

\begin{table}[h]
  \centering
  \caption{Comparison with state-of-the-art polar decoders.}
  \begin{tabular}{c c c c c}
    \toprule
                 & This work & \cite{Park2014} & \cite{Dizdar2014} & \cite{Sarkis2014}\\
    \midrule
    Dec. Algo.   & Fast-SSC  & BP              & SC                & Fast-SSC\\
    Code         & $(1024,512)$ & $(1024,512)$ & $(512,k)$         & $(1024,512)$\\
    IC Type      & FPGA      & ASIC            & ASIC              & FPGA\\
    Tech.        & 40 nm     & 65 nm           & 90 nm             & 40 nm\\
    $f$ (MHz)    & 231       & 300             & 6                 & 108\\
    Latency ($\mu$s) & 2.4   & 50              & 0.2               & 2\\
    T/P (Gbps)   & 237       & 4.7             & 2.9               & 0.5\\
    \bottomrule
  \end{tabular}
  \label{tab:cmp_results}
\end{table}

Recently, another fully unrolled polar decoder based on the less efficient SC algorithm has been presented in \cite{Dizdar2014}. That work is fully combinational with the exception of its input and output interfaces and as a result has a much lower frequency. The proposed decoder has a 14 times higher latency but is over 81 times faster than the 90 nm CMOS implementation of \cite{Dizdar2014}. The delay gain between 90 nm and 45 nm CMOS technology is 1.58 \cite{Wong2011}, still lower than the 3.4 to 4.6 factor between FPGA and ASIC. It should be noted that \cite{Dizdar2014} implemented a smaller polar code of length $N = 512$ instead of $N=1024$.

Table~\ref{tab:cmp_results} also presents results for a $(1024,512)$ polar code decoded using the implementation of \cite{Sarkis2014}. Our fully-unrolled, deeply-pipelined decoder has a throughput that is over 474 times greater than that previous Fast-SSC decoder implementation; while the latency is similar.

The proposed decoder has a throughput that is two orders of magnitude greater than that of state-of-the-art polar decoders.

\section{Conclusion}
In this Letter we presented a new architecture for a fully-unrolled,
deeply-pipelined polar decoder. We showed that a decoder for a $(1024,512)$
polar code implemented on an FPGA can achieve a coded throughput that is two
orders of magnitude faster than state-of-the-art polar decoders. At 237 Gbps, it is 51 to 81 times faster than the state-of-the-art ASIC implementations.

\section*{Acknowledgement}
Claude Thibeault is a member of ReSMiQ. Warren J. Gross is a member of ReSMiQ and SYTACom.

\bibliographystyle{IEEEtran}
\bibliography{IEEEabrv,letter.bib}

\end{document}